\documentclass[aps,prb,twocolumn,superscriptaddress,reprint]{revtex4-2}
\usepackage{float}
\usepackage{graphicx}
\usepackage{amsmath}
\usepackage{bm}
\usepackage{color}
\usepackage{amssymb}
\usepackage{mathrsfs}
\usepackage{dcolumn}
\usepackage{multirow}
\usepackage[colorlinks=true,linkcolor=blue,urlcolor=blue,citecolor=blue]{hyperref}

\begin{document}
	
\title{Orbital order in a bosonic $p$-band triangular lattice} 

\author{Hua Chen}
\email{Electronic address: hwachanphy@zjnu.edu.cn, he/him/his} 
\affiliation{Department of Physics, Zhejiang Normal University, Jinhua 321004, China}

\author{X. C. Xie}
\affiliation{International Center for Quantum Materials, School of Physics, Peking University, Beijing 100871, China}
\affiliation{Beijing Academy of Quantum Information Sciences, Beijing 100193, China}
\affiliation{CAS Center for Excellence in Topological Quantum Computation, University of Chinese Academy of Sciences, Beijing 100190, China}	
	
\begin{abstract}
We present a detailed study of the Bose-Hubbard model in a $p$-band triangular lattice by focusing on the evolution of orbital order across the superfluid-Mott insulator transition. Two distinct phases are found in the superfluid regime. One of these phases adiabatically connects the weak interacting limit. This phase is characterized by the intertwining of axial $p_\pm=p_x \pm ip_y$ and in-plane $p_\theta=\cos\theta p_x+\sin\theta p_y$ orbital orders, which break the time-reversal symmetry and lattice symmetries simultaneously. In addition, the calculated Bogoliubov excitation spectrum gaps the original Dirac points in the single-particle spectrum but exhibits emergent Dirac points. The other superfluid phase in close proximity to the Mott insulator with unit boson filling shows a detwined in-plane ferro-orbital order. Finally, an orbital exchange model is constructed for the Mott insulator phase. Its classical ground state has an emergent SO$(2)$ rotational symmetry in the in-plane orbital space and therefore enjoys an infinite degeneracy, which is ultimately lifted by the orbital fluctuation via the order by disorder mechanism. Our systematic analysis suggests that the in-plane ferro-orbital order in the Mott insulator phase agrees with and likely evolves from the latter superfluid phase.

\end{abstract}
	
\date{\today}
	
\maketitle

\section{Introduction}

Orbital order is a long-standing issue tracing back to the transition metal oxides~\cite{Tokura00,Maekawa04,Khomskii14}.
The precise mechanism driving orbital order remains largely unknown due to the intricate interplay among spin, orbital, charge, and lattice degrees of freedom in host crystals. In particular, a recent example of relevance is the nematic phase in iron-based superconductors, which entwines with spin Ising order, orbital order, and lattice structural distortion as dictated by symmetry~\cite{Fernandes14}. Among diverse theoretical proposals in addressing the origin of nematicity~\cite{Fradkin10,Fernandes19}, one interesting finding is that the orbital order in the nematic phase manifests its essential role in the metal-insulator transition and promotes an intermediate phase, {\it i.e.}, the orbital-selective Mott phase~\cite{Yu18}. This phase is characterized by the orbital-dependent Mott localization and interpolates the itinerant and Mott localized limits, validating the incipient Mott picture~\cite{Si09a,Si09b}. By contrast, a natural question may raise for bosonic systems: how the orbital order evolves in the superfluid-Mott insulator (SF-MI) transition.

Yet, much efforts have been denoted to the understanding of orbital order in electronic materials. While, the studies in bosonic systems are rare~\cite{Wu09,Li16}. Experimentally, artificial systems, such as ultracold atomic~\cite{Bloch07,Wirth11,Soltan12,Kock16,Zhou18} and photonic~\cite{Jacqmin14,Amo17,Amo19} systems, have been shown the exciting possibility of stimulating the crystals with $p$-orbital bosons in the first excited band. For instance, the Dirac points in the $p$-band hexagonal lattice are theoretically predicted by the early study~\cite{Wu08} and experimentally observed in photonic systems~\cite{Jacqmin14}. Later, the orbital edge state, which is extensively studied in graphene~\cite{Neto09,Novoselov11,Geim11}, is confirmed in the subsequent photonic experiment~\cite{Amo17}. More recently, the evidence of nematic superfluid (SF) phase in a hexagonal lattice, which is attributed to the orbital order, is also reported in ultracold atomic systems~\cite{Li20}.

The main purpose of our study is to give a comprehensive understanding of the orbital order in the SF-MI transition. The single-particle spectrum of the $p$-band triangular lattice exhibits a pair of Dirac points at the corners of hexagonal Brillouin zone (HBZ), resembling the low energy physics of graphene~\cite{Neto09,Novoselov11,Geim11}. The evolution of orbital order across the SF-MI transition is then studied based on the Bose-Hubbard model. In the weak-interacting limit, the $p$-band triangular lattice is frustrated due to the inability to simultaneously minimize both the kinetic and interacting energies. This weak-coupling SF phase is characterized by the intertwining of the axial $p_\pm=p_x\pm ip_y$ and in-plane $p_\theta=\cos\theta p_x+\sin\theta p_y$ orbital orders. Interestingly, the Bogoliubov excitation spectrum in this SF phase gaps the original Dirac points in the single-particle spectrum but exhibits emergent Dirac points. In the strong-interacting limit, the orbital order is also studied based on the orbital exchange model. We show that the classical ground state is of ferro-orbital type and enjoys an emergent SO$(2)$ rotational symmetry, which ensures an infinite degeneracy. The orbital fluctuation ultimately lifts the degeneracy and selects discrete quantum ground states through the order by disorder mechanism. Moreover, the phase diagram established by Gutzwiller approach interpolates these two limits. Besides these two phases, we find an intermediate SF phase with the ferro-orbital order. This intermediate phase survives in a wide range of low boson filling and gradually increase the occupation in the preferable in-plane orbital when approaching the Mott insulator (MI) with unit boson filling $n=1$. Our study provides strong clues that the ferro-orbital order in the MI $n=1$ phase likely evolves from the intermediate SF phase, facilitating the understanding on the role of orbital order in the  SF-MI transition. 

The reminder of this paper is organized as follow. In Sec.~\ref{sec:minimal}, we introduce the $p$-band tight-binding model in the triangular lattice as well as the Bose-Hubbard model. We establish the ground-state phase diagram by utilizing Gutzwiller approach in Sec.~\ref{sec:gutzwiller}. The orbital order in the weak-interacting limit is further studied with Bogoliubov approximation by treating the Bose-Hubbard interaction perturbatively in Sec~\ref{sec:weak}. In Sec.~\ref{sec:strong} the orbital exchange model is constructed to study the orbital order by treating the hopping processes as perturbations. Finally, we summarise and discuss the results in Sec.~\ref{sec:summary}.

\begin{figure}
	\centering
	\includegraphics[width=0.5\textwidth]{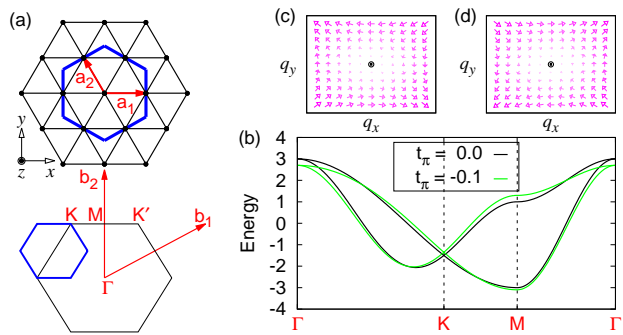}
	\caption{
		(a) The structure of triangular lattice and the hexagonal Brillouin zone. 
		The blue lines mark the Wigner-Seitz cell and the reduced Brillouin zone due to the umklapp scattering between the band minima at $M$ points. (b) The band structure of the tight-binding model in Eq.~(\ref{eq:TB}) with the $\sigma$-bonding $t_\sigma=1$ being the energy unit. The pseudovector fields $\bm{d}\equiv\left(d_z,d_x\right)$ near the Dirac point at $K$ (c) and $K^\prime$ (d) resemble the vortex in $XY$ systems with winding number $W=1$. 
	}
	\label{fig:band}
\end{figure}

\section{Minimal model}
\label{sec:minimal}

We begin with the tight-binding model that describes the hopping processes of bosons in the $p$-band triangular lattice depicted in Fig.~\ref{fig:band}(a). Introducing an orbital pseudospin representation, the momentum-space Hamiltonian in the basis $p_{\bm k}=\left[p_{x{\bm k}},p_{y{\bm k}}\right]^\text{T}$ reads
\begin{equation}
	\mathcal{H}_{\bm k} = 
	d_0\left({\bm k}\right)\tau_0+
	d_x\left({\bm k}\right)\tau_x+
	d_z\left({\bm k}\right)\tau_z,		
	\label{eq:TB}
\end{equation}
where $\tau_0$ and $\bm{\tau}$ are the identity matrix and Pauli matrices respectively, and the coefficients
$d_0\left(\bm{k}\right)=\left(t_\sigma+t_\pi\right)\sum_{i}\cos k_i$, and
$\{d_x\left(\bm{k}\right),d_z\left(\bm{k}\right)\}=\left(t_\sigma-t_\pi\right)/2\times\{\sqrt{3}\left(\cos k_3-\cos k_2\right),\cos k_1+\sum_{i}\cos k_i\}$. 
Here, the crystal momenta $\{k_1,k_2,k_3\}$ are measured along reciprocal lattice vectors $\{\bm{b}_1,\bm{b}_2,\bm{b}_3\equiv-\bm{b}_1-\bm{b}_2\}$, and the hopping integrals $t_\sigma$ and $t_\pi$ denote the $\sigma$ and $\pi$ bonding of $p$ orbitals, respectively. For the $\pi$ bonding, the bond vector lies in the nodal plane of $p$ orbitals. As a result, the strength of $\pi$ bonding is typically much weaker than that of $\sigma$ bonding. The band structure of the tight-binding model in Eq.~(\ref{eq:TB}) is plotted in Fig.~\ref{fig:band}(b). Notably, two bands cross at the Dirac points located at $K$ and $K^\prime$ points of HBZ. To describe the corresponding low-energy behavior around $K$ and $K^\prime$ points, we derive the effective $k\cdot p$ model  
\begin{eqnarray}
	\mathcal{H}_{K/K^\prime} \left({\bm q}\right) = 
	d_0\tau_0
	+d_x\tau_x
	+d_z\tau_z
	+\mathcal{O}\left(q^2\right)
	\label{eq:kp}
\end{eqnarray}
with the coefficients
\begin{eqnarray}
	d_0 = -\frac{3}{2}\left(t_\sigma+t_\pi\right),
	\{d_x,d_z\} = \pm\frac{3}{4}\sqrt{3}\left(t_\sigma-t_\pi\right)\{q_x,-q_y\}.\nonumber
\end{eqnarray}
Diagonalizing $\mathcal{H}_{K/K^\prime}\left({\bm q}\right)$ gives two non-interacting bands $E^\pm_{K/K^\prime}\left({\bm q}\right)=d_0\pm\sqrt{d_x^2+d_z^2}$, resulting in a linear dispersed Dirac point with the velocity $v=3\sqrt{3}/4(t_
\sigma-t_\pi)$. The pseudovector fields $\bm{d}\equiv\left(d_z,d_x\right)$ around $K$ and $K^\prime$ points, shown in Figs.~\ref{fig:band}(c) and \ref{fig:band}(d) respectively, have a $p$-wave symmetry. The topological charge of Dirac point is given by the winding number of pseudovector field: $W=\frac{1}{2\pi}\oint_\mathcal{C}\nabla\theta\left({\bm q}\right)\cdot d{\bm q}=1$, where $\theta\equiv\text{arctan}\left(d_x/d_z\right)$ and $\mathcal{C}$ is a contour enclosing the singular $K/K^\prime$ point, indicating that the Dirac point carries a $\pi$ Berry flux. The band minima are located at three inequivalent centres $M$ of HBZ edges, promoting a finite-momentum Bose-Einstein condensate for weakly interacting bosons. For non-interacting bosons, an infinite degenerate manifold of the single-particle ground state can be constructed by the linear superposition of the Bloch functions at these band minima. The umklapp scattering between the band minima transfers a lattice phonon which carries the momentum of multiple primitive reciprocal vectors. This process folds three $M$ points to $\Gamma$ point and underlies the reduced Brillouin zone (RBZ) and the enlarged Wigner-Seitz cell, as illustrated by the blue lines in Fig.~\ref{fig:band}(a).

\begin{figure}
	\centering
	\includegraphics[width=0.5\textwidth]{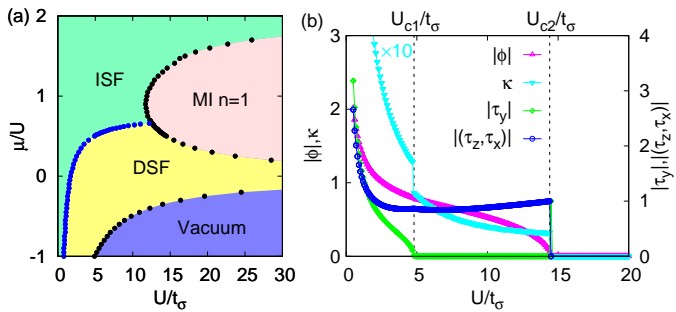}
	\caption{
		Gutzwiller approach.
		(a) The ground-state phase diagram in $\mu/U$ vs $U/t_\sigma$ plane accommodates three distinct phases, including (1) ISF, the superfluid phase intertwining axial and in-plane orbital orders; (2) DSF, the superfluid phase with the detwinned in-plane orbital order, and (3) MI $n=1$, the Mott insulator phase with unit filling $n=1$. The transition lines separating these three phases merge at a triple point $\left(U/t_\sigma,\mu/U\right)\approx\left(12.2,0.66\right)$. (b) Evolution of amplitude of condensate order parameter $\bm{\phi}=\left(\phi_x,\phi_y\right)$, compressibility $\kappa=\partial n/\partial \mu$, and orbital pseudospin $\bm{\tau}$ at fixed $\mu/U=0.5$. The black and blue dots in (a) mark the the critical points with the vanishing order parameters $|\bm{\phi}|$ and $|\tau_y|$, respectively. In the numerical calculations, the truncation of local Fock space $N_\pm=10$ for the maximum occupation in axial orbitals $p_\pm=p_x\pm ip_y$ and $t_\pi=0$ are used.
	}
	\label{fig:gutzwiller}
\end{figure}

Having established the single-particle physics, we are then in a position to study the effects of many-particle interactions. The interacting Hamiltonian can be generally constructed in terms of Haldane pseudopotentials by projecting a pair of particles into relative angular momenta, respecting the quantum statistics~\cite{Duncan90,chen18}. The Bose-Hubbard interaction which is mathematically described by zero relative angular momentum takes the form
\begin{equation}
	H_\text{I}=\frac{3}{2}U\sum_{i}\left[\hat{n}_{i}\left(\hat{n}_{i}-\frac{2}{3}\right)
	-\frac{1}{3}\hat{L}_{zi}^2\right],
\end{equation}
where $\hat{n}_{i}=\sum_{\alpha=x,y}p_{\alpha i}^\dagger p_{\alpha i}$ is the occupation operator and $\hat{L}_{zi}=-i\sum_{\alpha,\beta=x,y}\epsilon_{z\alpha\beta}p_{\alpha i}^\dagger p_{\beta i}$ is the $z$-component orbital angular momentum at $i$-th site~\cite{Isacsson05,Liu06,Umucalilar08}. Here $\epsilon_{\alpha\beta\gamma}$ is the Levi-Civita antisymmetric tensor. The interaction can be experimentally realized through the Feshbach resonance~\cite{Chin10} and optical nonlinearities~\cite{Carusotto13} for ultracold atomic and photonic systems, respectively.

\section{Gutzwiller approach}
\label{sec:gutzwiller}

To gain an overall understanding on the ground-state phase diagram, Gutzwiller approach~\cite{Gutzwiller63,Gutzwiller65,Kotliar91,Werner92} has its advance in capturing the physics in the intermediate regime of Hubbard interaction $U$, and straddles the limits of the weakly interacting SF and strongly interacting MI phases. It has been utilized to establish the phase diagram of $p$-band Bose-Hubbard model with different lattice geometries in the early study~\cite{Isacsson05}. This approach starts from the factorized local Fock state
\begin{eqnarray}
	\left|\Psi_\text{GW}\right\rangle=&&\prod_{i}\sum_{\text{F}}\eta_\text{F}^i\left|\text{F}\right\rangle_i, \nonumber\\
	\left|\text{F}\right\rangle_i=\frac{1}{\sqrt{n_+^\text{F}!n_-^\text{F}!}}&&
	\left(p_{+i}^{\dagger}\right)^{n_+^\text{F}} \left(p_{-i}^{\dagger}\right)^{n_-^\text{F}} \left|0\right\rangle \nonumber
\end{eqnarray}
where $n_{\pm}^\text{F}$ is the occupation of bosons in the axial orbitals $p_\pm=p_x\pm ip_y$ and $\eta_{\text{F}}^i$ is the probability weighting factor determined variationally. It takes into account that the multioccupation of bosons in the local orbitals are energetically costly. In numerical calculations, a truncation of the local Fock space is imposed and the filling of bosons is dictated by chemical potential $\mu$ for the grand canonical ensemble. We decompose the hopping terms in the tight-binding Hamiltonian as $p^\dagger_{\alpha i}p_{\beta j}\approx p^\dagger_{\alpha i}\phi_{\beta j}+\phi^*_{\alpha i}p_{\beta j}-\phi^*_{\alpha i}\phi_{\beta j}$ with the condensate order parameter $\phi_{\alpha i}=\sum_{\text{F}\text{F}^\prime} \eta_{\text{F}}^{i*}\eta_{\text{F}^\prime}^i{}_i\left\langle\text{F}|p_{\alpha i}|\text{F}^\prime\right\rangle_i$ entangling the local Fock states. The self-consistent solution of the order parameters $\phi_{\alpha i}$ requires an iterative minimization of the energy functional over the Wigner-Seitz cell in Fig.~\ref{fig:band}(a). The calculated phase diagram for $t_\pi=0$ shown in Fig.~\ref{fig:gutzwiller} (a) accommodates three different phases including two distinct SF phases and the MI phase. We have also verified that the phase diagram remains qualitatively robust against the perturbative $\pi$ bonding $t_\pi=-0.1t_\sigma$. To characterize the orbital order, we numerically evaluate the ground-state expectation of orbital pseudospin $\bm{\hat{\tau}}=\sum_{\alpha\beta} p^\dagger_{\alpha}\bm{\tau}_{\alpha\beta}p_{\beta}$. The axial orbital order $p_\pm=p_x\pm ip_y$ is characterized by the orbital polarization $\tau_y$ in $y$ axis, while the in-plane orbital order $p_\theta=\cos\theta p_x+\sin\theta p_y$ directing at angle $\theta$ with $x$ axis corresponds to the orbital polarization $\left(\tau_z,\tau_x\right)=\tau\left(\cos\left[2\theta\right],\sin\left[2\theta\right]\right)$ in $zx$ plane. Figure~\ref{fig:gutzwiller}(b) shows the detailed evolution of order parameters at fixed $\mu/U=0.5$, which determines the phases across the SF-MI transition. The stability of each phase is further checked with various sets of supercell sizes up to $8\bm{a}_1\times8\bm{a}_2$. As shown in Fig.~\ref{fig:gutzwiller}(b), two distinct SF phases share a non-vanishing uniform order parameter $|\bm{\phi}_i|$ and can be distinguished by the ground-state expectation value of orbital pseudospin $\hat{\bm{\tau}}$. Initially, the ground state at weak Hubbard interaction $U$ develops an intertwined order by entangling both the axial and in-plane orbital orders. The former is characterized by the alternating signs in adjacent rows but an identical amplitude of $\tau_y^i$ therefore suggesting the ordering of antiferro-orbital angular momentum, while the latter is indicated by the uniform $\left(\tau^i_z,\tau^i_x\right)$ implying the ferro-orbital order. Therefore, this superfluid phase intertwining axial and in-plane orbital orders is denoted as ISF. The detailed pattern of orbital orders will be further discussed in Sec.~\ref{sec:weak}. At the critical Hubbard interaction $U_{c1}$, the orbital pseudospin is completely aligned in the $zx$ plane $\left(\tau_z,\tau_x\right)$, showing a detwinned ferro-orbital order. This superfluid phase is thus denoted as DSF. Since the inter-site hopping process is treated at the mean-filed level in Gutzwiller approach, we will show in Sec.~\ref{sec:strong} that the orientation of in-plane orbital is solely determined by the quantum fluctuation due to the orbital anisotropy. With further increasing Hubbard interaction $U>U_{c2}$, the MI phase obtained within Gutzwiller approximation is simply a product of local Fock states and is thus featureless. As shown in Fig.~\ref{fig:gutzwiller}(b), the ISF-DSF-MI transition driven by he Hubbard interaction $U$ is well detected by the discontinuous jumps of the compressibility $\kappa=\partial n/\partial\mu$. Therefore, the phase transition discussed here may be experimentally probed by measuring the boson filling $n$. Below we shall justify the orbital orders above from two extrema limits.

\begin{figure}
	\centering
	\includegraphics[width=0.5\textwidth]{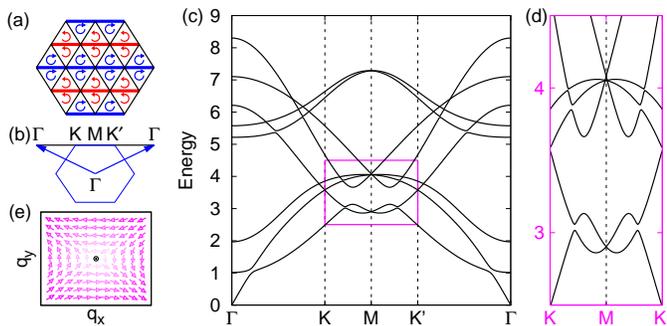}
	\caption{ Bogoliubov approach.
		(a) The condensed ground state intertwines the ferro-orbital order with the in-plane orbital being parallel to the bond, and the antiferro-orbital angular momentum with the spatial pattern indicated by red and blue lines, supporting staggered fluxes. (c) The Bogoliubov excitation spectrum along the high symmetry line indicated in (b). (d) The detailed Bogoliubov excitation spectrum indicated by the pink box in (c) exhibits Dirac bosons.	 
		(e) The pseudovector field $\bm{d}$ of Dirac boson in the lowest two excitation branches around $M$ point has winding number $W=1$. The parameters used in numerics are $\left(t_\sigma,t_\pi,nU\right)=\left(1,0,1\right)$.
	}
	\label{fig:bogoliubov}
\end{figure}

\section{Weak-coupling approach}
\label{sec:weak}

In the weakly interacting limit, the Hubbard interaction $U$ is treated perturbatively. The operators can be decomposed in terms of quantum fluctuations $\tilde{p}_{\alpha\bm{k}\ell}$ around the condensed ground-state wave function $\phi_{\alpha \ell}$
\begin{equation}
	p_{\alpha\bm{k}\ell} = \phi_{\alpha \ell}\delta\left(\bm{k}\right)+\tilde{p}_{\alpha\bm{k}\ell}\text{, }
	\alpha=x,y  
\end{equation}
with $\ell$ specifying the sublattice in the Wigner-Seitz cell.
In the spirit of Bogoliubov approximation~\cite{Bogolyubov47,Abrikosov63}, the Hamiltonian is expanded in powers of quantum fluctuations and is truncated up to the quadratic order. 
Detailed derivations are presented in Appendix~\ref{app:Bgl}.
The zeroth-order terms in this expansion determine the energy functional $\varepsilon\left(\bm{\phi}^*,\bm{\phi}\right)$ of the condensate at $\Gamma$ point in RBZ. The time-dependent Gross-Pitaevskii equation can be readily derived via the Euler-Lagrange equation
\begin{equation}
	\frac{\partial\mathcal{L}}{\partial\phi_{\alpha\ell}^*}
	-\frac{d}{dt}\left(\frac{\partial\mathcal{L}}{\partial\dot{\phi}_{\alpha\ell}^*}\right)=0, 
\end{equation}
where the Lagrangian $\mathcal{L}=\sum_{\alpha\ell}i\hbar\left(\phi_{\alpha \ell}^*\dot{\phi}_{\alpha\ell}-\phi_{\alpha \ell}\dot{\phi}_{\alpha\ell}^*\right)-\varepsilon\left(\bm{\phi}^*,\bm{\phi}\right)$~\cite{Pethick08}. The ground state can be numerically solved through the imaginary-time evolution of Gross-Pitaevskii equation by propagating an initial trial state~\cite{Dalfovo99}. Mathematically, this procedure is equivalent to the minimization of the energy functional $\varepsilon\left(\bm{\phi}^*,\bm{\phi}\right)$, which causes the linear order terms in $\tilde{p}_{\alpha\bm{k}\ell}$ to vanish. The calculated ground-state condensate develops an intertwined order, confirming the results from Gutzwiller approach. The ferro-orbital order is characterized by orientating the in-plane orbital parallel to the bond, which breaks the lattice rotational symmetry. The antiferro-orbital angular momentum is characterized by the alternating sign of $\tau_y$ along the direction perpendicular to the in-plane orbital, and breaks the time-reversal symmetry as well as the lattice translational symmetry. Interestingly, the symmetry breaking of this weak-coupling phase, which has been studied in details, is shown to be universal in the strong-coupling regime with boson filling $n\ge2$~\cite{Wu06}. As schematically depicted in Fig.~\ref{fig:bogoliubov}(a), the staggered flux pattern of the ground-state condensate arising from the time-reversal symmetry breaking is characterized by the bond current $J_{ij}=-i\sum_{\alpha\beta}t_{\gamma}^{\alpha\beta}(\langle p^\dagger_{\alpha i}p_{\beta j}\rangle-\text{c.c.})\delta_{j,i\pm \bm{a}_\gamma}$ where the hopping matrix $t_{\gamma}=[(t_\sigma+t_\pi)\tau_0+(t_\sigma-t_\pi)(\cos[2\theta_\gamma]\tau_z+\sin[2\theta_\gamma]\tau_x)]/2$ and $\theta_\gamma$ is the azimuthal angle of $\bm{a}_\gamma$. The early studies find pure axial orbital orders with different lattice geometries, which support bond currents as a natural consequence~\cite{Liu06,Xu16,Liberto16}. In contrast, the intertwined orbital order in the present study, due to the inability to simultaneously minimize both kinetic and interacting energies, originates from the geometric frustration of the triangular lattice. Having settled the ground state, we then proceed with the quadratic order 
\begin{equation}
	\mathcal{H}^{(2)}_{\bm{k}}=
	\frac{1}{2}
	\left[\tilde{\bm{p}}^\dagger_{\bm{k}},\tilde{\bm{p}}_{-\bm{k}}\right]
	\left[
	\begin{matrix}
		X_{\bm{k}} & Y \\
		Y^\dagger  & X_{-\bm{k}}
	\end{matrix}
	\right]
	\left[
		\begin{matrix}
		\tilde{\bm{p}}_{\bm{k}} \\
		\tilde{\bm{p}}^\dagger_{-\bm{k}}  
	\end{matrix}
	\right],
	\label{eq:bogoliubov}
\end{equation}
which describes the Bogoliubov excitation on top of the ground-state condensate. The diagonal terms $X_{\pm\bm{k}}$ in Eq.~(\ref{eq:bogoliubov}) receive contributions from both the hopping processes and the self-energy correction of Hubbard interaction $U$, while the off-diagonal terms $Y$ ($Y^\dagger$) describe the anomalous processes in which a pair of bosons scatter with each other into the excited states (condensates). As shown in Fig.~\ref{fig:bogoliubov}(c), the Bogoliubov excitation spectrum has a gapless Goldstone mode around $\Gamma$ point in RBZ as the signature of $U(1)$ symmetry breaking. 
Interestingly, we find that the Bogoliubov spectrum along the high symmetry line $K$-$K^\prime$ gaps out the original Dirac points at $K$ and $K^\prime$ points in the band structure but exhibits emergent Dirac bosons at $M$ point. As a representative example, the Dirac point in the lowest two branches is identified by numerically calculating the winding number shown in Fig.~\ref{fig:bogoliubov}(e). It is worth mentioning that the Dirac bosons only exist on one edge of RBZ parallel to the bond that is selected by the in-plane orbital order, serving as a fingerprint of lattice rotational symmetry breaking. We have also checked that the Dirac related physics is robust for $t_\pi=-0.1t_\sigma$. While the early study focuses on the Dirac fermion in the band structure with staggered fluxes~\cite{Wang17}, the present study investigates the Dirac bosons in the elementary excitation on top of the Bose-Einstein condensate with staggered fluxes instead.

\begin{figure}
	\centering
	\includegraphics[width=0.5\textwidth]{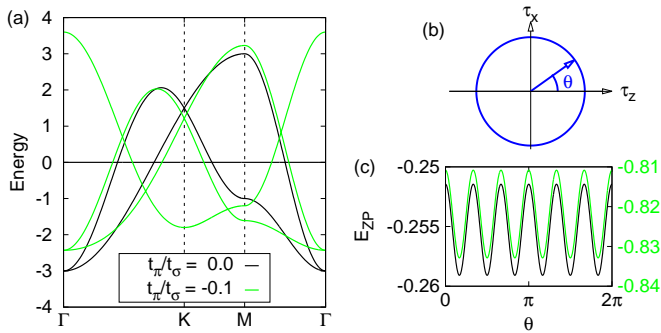}
	\caption{ Strong-coupling approach.
		(a) The eigenvalues of orbital interaction matrix $\Lambda_{\bm{k}}$ within the classical approximation.
		(b) The emergent SO$(2)$ rotational symmetry in the easy plane $(\tau_z,\tau_x)$ of orbital pseudospin space.
		(c) The zero-point energy $E_\text{ZP}\left(\theta\right)$ in Eq.~(\ref{eq:zp}) from the correction of orbital fluctuations has minima at $\theta=\pi/6+\mathbb{Z}\pi/3$. The energies in (a) and (c) are in units of $t_\sigma^2/16U$.
	}
	\label{fig:sw}
\end{figure}
 
\section{Strong-coupling approach}
\label{sec:strong}

Finally, we turn to the strong-coupling limit in which the virtual hopping processes are treated perturbatively. Since the charge excitation in MI phase is suppressed by the charge gap proportional to Hubbard interaction $U$, the orbital fluctuation is the remaining low energy degree of freedom. Following the standard second-order perturbation theory, the effective low-energy physics for the MI $n=1$ phase is captured by the following orbital exchange model
\begin{eqnarray}
	H_\text{OE}=J\sum_{\langle ij \rangle\parallel\bm{a}_\gamma}\tau^i_\gamma\tau^j_\gamma
	+J^\prime \sum_{\langle ij \rangle}
	\left(\bm{\tau}^i\cdot\bm{\tau}^j+2\tau^i_y\tau^j_y\right)
	\label{eq:orbex}
\end{eqnarray}
with
\begin{eqnarray}
	\tau_\gamma=\tau_z\cos\left[2\theta_\gamma\right]+\tau_x\sin\left[2\theta_\gamma\right]. \nonumber
\end{eqnarray}
Detailed derivations are presented in Appendix~\ref{app:OEX}.
The ferro-orbital exchange $J=-(t_\sigma-t_\pi)^2/16U$ in Eq.~(\ref{eq:orbex}) is inherently anisotropic originating from the anisotropic shape of $p$ orbitals. In contrast, the exchange $J^\prime=-t_\sigma t_\pi/8U$ is antiferro-orbital due to the opposite sign of $t_\sigma$ and $t_\pi$. The first term in Eq.~(\ref{eq:orbex}) involving interacting orbital degrees of freedom is coined as the compass model~\cite{Kugel82,Nussinov15}. The ground state is first studied by treating the orbital pseudospin $\bm{\tau}$ as a classical vector. The orbital interaction can be minimized via the diagonalization of the orbital exchange Hamiltonian in momentum space $H_\text{OE} = \sum_{\bm{k}}\bm{\tau}^{-\bm{k}}\Lambda_{\bm{k}}\bm{\tau}^{\bm{k}}$. As shown in Fig.~\ref{fig:sw}(a), the lowest eigenvalue of $\Lambda_{\bm{k}}$ is found at $\Gamma$ point in RBZ and has a twofold degeneracy, suggesting that the classical ground state is ferro-orbital ordering. The degenerated eigenvalue has important implications on the structure of orbital order. A close inspection on the Hamiltonian in Eq.~(\ref{eq:orbex}) reveals that the $y$ component of orbital pseudospin $\tau_y$ is decoupled from the other two components $\tau_{z,x}$, in which the lowest degenerate eigenvalue arises. It indicates that the ordering of orbital pseudospin $\bm{\tau}$ lies in the $zx$ plane. More importantly, this degeneracy renders a continuous $\text{SO}(2)$ rotational symmetry of orbital pseudospin $(\tau_z,\tau_x)=\tau(\cos\theta,\sin\theta)$ as shown in Fig.~\ref{fig:sw}(b). Note that this symmetry restricted to the classical ground state is emergent and is not an exact symmetry of the orbital exchange model, which is only invariant under finite point group rotations. The orbital order of the classical ground state evolves in the $zx$ plane without any energy cost, which makes the system particularly susceptible to quantum fluctuations. Following Holstein-Primakoff spin wave theory~\cite{Holstein40}, the zero-point energy arises from the correction of quantum fluctuations, and is studied as a function of the rotation $\theta$ about the $y$ axis of orbital pseudospin. To the leading order, the zero-point energy takes the form
\begin{equation}
	E_\text{ZP}(\theta)=\frac{1}{2N}\sum_{\bm{k}}\omega_{\bm{k}}\left(\theta\right)+6J+12J^\prime,
	\label{eq:zp}
\end{equation}
where $N$ is the number of sites, the orbital excitation $\omega_{\bm{k}}(\theta)=2\sqrt{[\varphi_{\bm{k}}(\theta)+2\varphi^\prime_{\bm{k}}-6J-12J^\prime]^2-[\varphi_{\bm{k}}(\theta)+\varphi_{\bm{k}}^\prime]^2}$, and the auxiliary functions $\{\varphi_{\bm{k}}(\theta),\varphi_{\bm{k}}^\prime\}=\{2J\sum_{\gamma}\sin^2[2\theta_\gamma+\theta],4J^\prime\}\cos[\bm{k}\cdot\bm{a}_\gamma]$. Detailed derivations are presented in Appendix~\ref{app:LOW}. The numerical evaluation of zero-point energy is shown in Fig.~\ref{fig:sw}(c). The orbital fluctuation lifts the degeneracy protected by the continuous $\text{SO}(2)$ rotational symmetry, and selects the quantum ground state at $\theta=\pi/6+\mathbb{Z}\pi/3$ ($\mathbb{Z}$ is an integer). This mechanism is known as order by disorder in frustrated spin systems~\cite{Villain80,Henley89,Lacroix11,Diep13,Green18}. 
 
\section{Conclusion and discussion}
\label{sec:summary}

To summarize, we have studied the evolution of orbital ordering across the SF-MI transition in the $p$-band triangular lattice.  The ground-state phase diagram is first established by Gutzwiller approach, which interpolates continuously between two extreme limits, deep in SF phase and deep in MI phase. The orbital orders in these two limits are further examined by the perturbation approaches. With systematic analyses, we identify an intermediate SF phase with the detwined in-plane ferro-orbital order, which correctly reproduce the orbital order in the MI $n=1$ phase. It is worth remarking several directions for further studies. The quantum fluctuations, which can be partially restored with the cluster Gutzwiller approach~\cite{Luhmann13,Bai18}, may deserve to be studied for its role in selecting the orbital order in the vicinity of SF-MI transition. Alternatively, it is also interesting to investigate the details of SF-MI transition within a single unified method, {\it e.g.} the quantum Monte-Carlo simulation. Finally, we close by briefly discussing the dissipation. The experimental realization of the Bose-Hubbard model in photonic systems involves light-matter interactions, which may be better described as an open system. Therefore, another direction to generalize our work is to study the effect of dissipation.

\section*{Acknowlegdgement}

We thank Congjun Wu and W. Vincent Liu for helpful discussions. 
This work is supported by the National Natural Science Foundation of China under Grants No. 11704338, No. 11534001, No. 11504008, and the National Basic Research Program of China under Grant No. 2015CB921102. 

\appendix

\section{Details of Bogoliubov approach}
\label{app:Bgl}

In the weak-coupling limit, the band minima are occupied by a macroscopic number of bosons $N_0$ at zero crystal momentum $\bm{k}=0$ in RBZ, which can be expressed in terms of the ground-state wave function $N_0=\sum_{\alpha\ell}|\phi_{\alpha \ell}|^2$. Here $\alpha$ and $\ell$ specify the orbital and sublattice degree of freedom, respectively. The quantum effects arising from the commutation relation $[p_{\alpha\ell\bm{k}=0},p^\dagger_{\alpha^\prime\ell^\prime\bm{k}=0}]=
\delta_{\alpha,\alpha^\prime}\delta_{\ell,\ell^\prime}$ are suppressed by the macroscopic occupation $N_0$. The occupation of the excited states at non-zero momentum $\bm{k}\ne0$ with the corresponding operator $p_{\alpha\ell\bm{k}\ne0}$ is generally small and is treated as quantum fluctuations $\tilde{p}_{\alpha\ell\bm{k}}$. The Hamiltonian can be reexpressed by splitting the operators into the ground-state wave function $\phi_{\alpha \ell}$ and the quantum fluctuations $\tilde{p}_{\alpha\ell\bm{k}}$. To the lowest order in $\tilde{p}_{\alpha\ell\bm{k}}$, one neglect all fluctuations and obtain the energy functional per unit cell
\begin{eqnarray}
 \varepsilon\left(\bm{\phi}^*,\bm{\phi}\right) &=&
 \mathcal{T}_{\Gamma}\left(\bm{\phi}^*,\bm{\phi}\right)
 +\mathcal{I}\left(\bm{\phi}^*,\bm{\phi}\right)
  \label{eq:efs}
 \end{eqnarray}
with the energy functional 
\begin{eqnarray}
	\mathcal{T}_{\bm{k}}\left(\bm{\phi}^*,\bm{\phi}\right)
	= \frac{1}{N_\text{uc}}\sum_{\alpha\alpha^\prime}\sum_{\ell\ell^\prime}
	t^{\alpha\alpha^\prime}_{\ell\ell^\prime}(\bm{k})
	\phi_{\alpha \ell}^*\phi_{\alpha^\prime \ell^\prime}
\end{eqnarray}
and
 \begin{eqnarray}
 \mathcal{I}\left(\bm{\phi}^*,\bm{\phi}\right)
 &=&\frac{3U}{2N_\text{uc}^2}\sum_{\ell}
 \left[\right. 
 |\phi_{x \ell}|^4+|\phi_{y \ell}|^4 
 +\frac{4}{3}|\phi_{x \ell}|^2|\phi_{y \ell}|^2 \nonumber\\
 &+&\frac{1}{3}\phi_{x \ell}^{*2}\phi_{y \ell}^2
 +\frac{1}{3}\phi_{x \ell}^2\phi_{y \ell}^{*2}
 \left.\right]
\end{eqnarray}
arising from the hopping processes and the Hubbard interaction, respectively.
Here $N_\text{uc}$ is the number of unit cells and $t^{\alpha\alpha^\prime}_{\ell\ell^\prime}(\bm{k})$ is the matrix elements describing the process that the bosons with crystal momentum $\bm{k}$ hop among the sublattices in the Wigner-Seitz cell. The ground-state wave function is obtained by minimizing the energy functional in Eq.~(\ref{eq:efs}), which ensures the linear order terms in $\tilde{\bm{p}}_{\alpha\ell\bm{k}}$ to vanish. The quadratic order terms can be written as 
\begin{eqnarray}
	\mathcal{H}^{(2)}_{\bm{k}}=
	\frac{1}{2}
	\left[\tilde{\bm{p}}^\dagger_{\bm{k}},\tilde{\bm{p}}_{-\bm{k}}\right]
	\left[
	\begin{matrix}
		X_{\bm{k}} & Y \\
		Y^\dagger  & X_{-\bm{k}}
	\end{matrix}
	\right]
	\left[
	\begin{matrix}
		\tilde{\bm{p}}_{\bm{k}} \\
		\tilde{\bm{p}}^\dagger_{-\bm{k}}  
	\end{matrix}
	\right].
\end{eqnarray}
The elements of diagonal matrix in the orbital-sublattice basis $\tilde{\bm{p}}_{\bm{k}}=\left[
\tilde{p}_{x1\bm{k}},
\tilde{p}_{x2\bm{k}},
\tilde{p}_{x3\bm{k}},
\tilde{p}_{x4\bm{k}},
\tilde{p}_{y1\bm{k}},
\tilde{p}_{y2\bm{k}},
\tilde{p}_{y3\bm{k}},
\tilde{p}_{y4\bm{k}}
\right]$ are 
\begin{eqnarray}
	X_{\bm{k}}^{\alpha\ell,\alpha^\prime\ell^\prime}
	&=& -(\mathcal{T}_\Gamma+2\mathcal{I})\delta_{\alpha,\alpha^\prime}
	\delta_{\ell,\ell^\prime}
	+t^{\alpha\alpha^\prime}_{\ell\ell^\prime}(\bm{k})\nonumber\\
	&+& \frac{2U}{N_\text{uc}}
	\delta_{\alpha,\alpha^\prime}
	\delta_{\ell,\ell^\prime}
	[|\phi_{x\ell}|^2+|\phi_{y\ell}|^2+2|\phi_{\alpha\ell}|^2]
	\nonumber\\
	&+& \frac{2U}{N_\text{uc}}	
	(1-\delta_{\alpha,\alpha^\prime})	
	\delta_{\ell,\ell^\prime}
	[\phi_{x\ell}^*\phi_{y\ell}+\phi_{y\ell}^*\phi_{x\ell}].
	\label{eq:H2X}
\end{eqnarray} 
It is worth mentioning that the first term in Eq.~(\ref{eq:H2X}) arises from the conservation of bosons.
The total bosons consist of condensed bosons at zero momentum and excited bosons in the fluctuating fields $N=N_0+\sum_{\alpha\ell\bm{k}}\tilde{p}_{\alpha\ell\bm{k}}^\dagger\tilde{p}_{\alpha\ell\bm{k}}$.
The elements of off-diagonal matrix are as follow
\begin{eqnarray}
	Y^{\alpha\ell,\alpha^\prime\ell^\prime}&=&\frac{U}{N_\text{uc}}
	\delta_{\alpha,\alpha^\prime}
	\delta_{\ell,\ell^\prime}
	[\phi_{x \ell}^2+\phi_{y \ell}^2+2\phi_{\alpha \ell}^2]\nonumber\\
	&+&\frac{2U}{N_\text{uc}}
	(1-\delta_{\alpha,\alpha^\prime})	
	\delta_{\ell,\ell^\prime}
	\phi_{x \ell}\phi_{y \ell}.
\end{eqnarray}

The Bogoliubov excitation is determined by the eigenvalues of the Bogoliubov dispersion matrix~\cite{BLAIZOT86}
\begin{eqnarray}
	\bm{\sigma}_z
	\left[
	\begin{matrix}
		X_{\bm{k}} & Y \\
		Y^\dagger  & X_{-\bm{k}}
	\end{matrix}
	\right]
	=
	\left[
	\begin{matrix}
		X_{\bm{k}} & Y \\
		-Y^\dagger  & -X_{-\bm{k}}
	\end{matrix}
	\right],
\end{eqnarray}
where $\bm{\sigma}_z=\text{diag}\left(+\bm{1},-\bm{1}\right)$ is the block Pauli matrix.

\section{Derivation of orbital exchange Hamiltonian}
\label{app:OEX}
\begin{table*}
	\caption{\label{tab:gamma-s} 
		The $i$-th eigenstate $\Gamma_n^i$ of the Hamiltonian $H_\text{I}$ in Eq.~(\ref{eq:HIS}) with the corresponding eigenenergy $E_{\Gamma_n^i}$ for $p^{n=0,1,2}$ configurations.}
	\begin{ruledtabular}
		\begin{tabular}{cccccccc}
			&\multicolumn{1}{c}{$p^0$ configuration}&\multicolumn{2}{c}{$p^1$ configuration}&\multicolumn{3}{c}{$p^2$ configuration}\\ \hline
			$i$	
			&	$1$	
			&	$1$	&	$2$
			&	$1$	&	$2$	& 	$3$	\\ 
			$E_{\Gamma_n^i}$	
			&	$0$	
			&	$0$	&	$0$
			&	$2U$	&	$4U$	& 	$2U$	\\ 	
			$\left|n_+n_-\right\rangle$	
			&	$\left|0,0\right\rangle$	
			&	$\left|1,0\right\rangle$	&		$\left|0,1\right\rangle$
			&	$\left|2,0\right\rangle$	&	    $\left|1,1\right\rangle$	&		$\left|0,2\right\rangle$		\\					
		\end{tabular}
	\end{ruledtabular}
\end{table*}

To derive the effective low-energy Hamiltonian, we shall first diagonalize the on-site Hubbard interaction 
\begin{eqnarray}
	H_\text{I}&=&\frac{3}{2}U\left[\hat{n}\left(\hat{n}-\frac{2}{3}\right)
-\frac{1}{3}\hat{L}_{z}^2\right].
	\label{eq:HIS}
\end{eqnarray}
It is easy to show that the Hamiltonian in Eq.~(\ref{eq:HIS}) commutes with both the total occupation operator $\hat{n}$ and the $z$-component angular momentum $\hat{L}_z$. Moreover, the matrix representation of the operators $\hat{n}=\hat{n}_+ + \hat{n}_-$ and $\hat{L}_z=\hat{n}_+ - \hat{n}_-$ is diagonal in the basis of axial orbitals $p_{\pm}=p_x\pm ip_y$. Here, $\hat{n}_\pm$ correspond to the occupation operators of axial orbitals $p_\pm$. Therefore, the eigenstate of the Hamiltonian in Eq.~(\ref{eq:HIS}) can be labelled by the corresponding quantum number
\begin{eqnarray}
 \left|n_{+}n_{-}\right\rangle=
 \frac{1}{\sqrt{n_+!n_-!}}\left(p_+^\dagger\right)^{n_+}\left(p_-^\dagger\right)^{n_-}\left|0,0\right\rangle.	
\end{eqnarray}
The orbital exchange model in the MI $n=1$ phase involves the $p^{n=0,1,2}$ configurations.
The eigenstates $\Gamma_n^i$ with the corresponding eigenenergies $E_{\Gamma_n^i}$ are listed in Table.~\ref{tab:gamma-s}. 
The $p^1$ configuration with zero energy is an orbital doublet with one boson occupying either $p_+$ or $p_-$ orbital.
Note that the charge excitation $\left(p^1\right)_i\left(p^1\right)_j\rightleftharpoons\left(p^2\right)_i\left(p^0\right)_j$ through the hopping processes $t_{\alpha\beta}d^\dagger_{i\alpha}d_{j\beta}$ has an energy gap that is proportional to the Hubbard interaction $U$. 
In the large-$U$ limit, the effective low-energy model is described by the second-order hopping process with both the initial and final states in $p^{1}$ configuration, which involves no charge gap. Let us first derive the orbital exchange interaction along the $\bm{a}_1=\hat{\bm{x}}$ bonds.
Employing the second-order perturbation theory~\cite{Kuklov03}, the matrix form of orbital exchange interaction is given by
\begin{eqnarray}
	\mathbb{J}_{kl,k^\prime l^\prime}=
	&-&\sum_{mn}\frac{1}{E_{\Gamma_{2}^m}+E_{\Gamma_0^n}} \nonumber\\
	&\times&\left\langle
	\underset{i\text{-th site}}{\Gamma_1^k}
	\underset{j\text{-th site}}{\Gamma_1^l}\right|
	\sum_{\alpha\beta}t_{\alpha\beta}^*p^\dagger_{j\beta}p_{i\alpha}
	\left| \underset{i\text{-th site}}{\Gamma_2^m}
	\underset{j\text{-th site}}{\Gamma_0^n} 
	\right\rangle \nonumber\\
	&\times&\left\langle
	\underset{i\text{-th site}}{\Gamma_2^m}
	\underset{j\text{-th site}}{\Gamma_0^n}\right|
	\sum_{\delta\gamma}t_{\delta\gamma}p_{i\alpha}^\dagger p_{j\gamma}
	\left| \underset{i\text{-th site}}{\Gamma_1^{k^\prime}}
	\underset{j\text{-th site}}{\Gamma_1^{l^\prime}}
	\right\rangle \nonumber\\
	&+&{i\leftrightarrow j}. 
\end{eqnarray}
To describe the orbital exchange Hamiltonian, we introduce the orbital pseudospin operators  
$\{\tau_+,\tau_-\} \equiv \{p^\dagger_{x}p_{y},p^\dagger_{y}p_{x}\}$,
which flip the states of the orbital doublet in $p^1$ configuration. The $z$ component of pseudospin $\bm{\tau}$-vector follows through the spin-$1/2$ angular momentum algebra $\tau_z=\left[\tau_+,\tau_-\right]$. 
A lengthy but straightforward algebra leads the orbital exchange Hamiltonian along the $\bm{a}_1$ bonds
\begin{equation}
	H_\text{OE}^{1} = \sum_{\langle ij\rangle} 
	\left[
	J\tau_z^i\tau_z^j 
	+ J^\prime\left(\bm{\tau}^i\cdot\bm{\tau}^j+2\tau^i_y\tau^j_y\right)
	\right]
\end{equation}
with 
\begin{eqnarray}
	\{J,J^\prime\}&=&\{-\frac{\left(t_\sigma-t_\pi\right)^2}{16U},\frac{t_\sigma t_\pi}{8U}\}. 
\end{eqnarray}
Having derived the orbital exchange interaction $H_\text{OE}^{1}$ along bond vector $\bm{a}_1$, the interaction $H_\text{OE}^{2,3}$ has exactly the same form with $H_\text{OE}^{1}$ if the orbital pseudospin operators $\bm{\tau}$ are defined in the local coordinate. Thus, the connection between the local and global coordinates (the global $x$ axis along $\bm{a}_1$ bond vector) is linked by a rotation of $\theta=\frac{2\pi}{3},\frac{4\pi}{3}$ about $z$ axis,
corresponding to the $\bm{a}_2,\bm{a}_3$ bonds, respectively. Under this rotation, the $p$ orbital wave functions transform as 
\begin{subequations}
	\begin{eqnarray}
		p_{x} &\to& \cos\theta p_{x} - \sin\theta p_{y}, \\
		p_{y} &\to& \sin\theta p_{x} + \cos\theta p_{y}.
	\end{eqnarray} 	
\end{subequations}
Accordingly, the pseudospin operators $\bm{\tau}$ transform as follow
\begin{subequations}
	\begin{eqnarray}
		\tau_z &\to& \sin\left[2\theta\right]\tau_x+\cos\left[2\theta\right]\tau_z, \\
		\tau_x &\to& \cos\left[2\theta\right]\tau_x-\sin\left[2\theta\right]\tau_z, \\
		\tau_y &\to& \tau_y. 
	\end{eqnarray}	
\end{subequations}
The pseudospin vector $\bm{\tau}$ is rotated by $2\theta$ about its $y$ axis in the pseudospin space.
It is now straightforward to obtain the Hamiltonian $H_\text{OE}^{2,3}$ by replacing the pseudospin $\bm{\tau}$ in $H_\text{OE}^{1}$. 
Finally, the total orbital exchange Hamiltonian takes the form 
\begin{eqnarray}
	H_\text{OE}=J\sum_{\langle ij \rangle\parallel\bm{a}_\gamma}\tau^i_\gamma\tau^{j}_\gamma
	+J^\prime \sum_{\langle ij \rangle}
	\left(\bm{\tau}^i\cdot\bm{\tau}^j+2\tau^i_y\tau^j_y\right)
	\label{eq:OES}
\end{eqnarray}
with
\begin{subequations}
	\begin{eqnarray}
		\tau_\gamma=\tau_z\cos\left[2\theta_\gamma\right]+\tau_x\sin\left[2\theta_\gamma\right],
		\{\theta_1,\theta_2,\theta_3\}=\{0,\frac{2}{3}\pi,\frac{4}{3}\pi\}. \nonumber
	\end{eqnarray}
\end{subequations}
Unlike spin systems, the orbital exchange is anisotropic. It roots in the fact that the hopping processes are bond dependent due to the spatial orientation of $p$ orbitals.

\section{Linear orbital wave}
\label{app:LOW}
Having established the classical ground state of the orbital exchange Hamiltonian $H_\text{OE}$ in Eq.~(\ref{eq:OES}), 
we then proceed to derive the Hamiltonian that describes the orbital excitation.
The orbital pseudospin operators, obeying the angular momentum algebra of spin $T=1/2$, can be expressed in terms of Holstein-Primakoff bosons~\cite{Holstein40}
\begin{subequations}
\begin{eqnarray}
	\tau_x&=&\sqrt{2T}\left(a^\dagger+a\right), \\
	\tau_y&=&i\sqrt{2T}\left(a^\dagger-a\right), \\	
	\tau_z&=&2\left(T-a^\dagger a\right).
\end{eqnarray}	
\end{subequations}
Note that the classical ground state enjoys a continuous SO$(2)$ rotational symmetry, 
which transforms the orbital pseudospin as
\begin{subequations}
	\begin{eqnarray}
		\tau_z&\to& \cos\theta\tau_z+\sin\theta\tau_x,\\
		\tau_x&\to&-\sin\theta\tau_z+\cos\theta\tau_x.
	\end{eqnarray}
\end{subequations} 
The pseudospin in Eq.~(\ref{eq:OES}) is first replaced by the above transformation.
Expanding in powers of $T$ followed by Fourier transformation then leads to the following $\theta$-dependent Hamiltonian 
\begin{eqnarray}
	H_\text{OE}(\theta)=
	4T^2NE_\text{c} 
	+2T\sum_{\bm{k}}\hat{\mathcal{H}}_\text{LOW}\left(\bm{k},\theta\right)
	+\mathcal{O}(\sqrt{T})\hspace{8mm}
	\label{eq:HOES}
\end{eqnarray}
The  first term $E_\text{c}=3J+6J^\prime$ in Eq.~(\ref{eq:HOES}) recovers the classical ground energy per site. 
The second term in Eq.~(\ref{eq:HOES}) describes the linear orbital wave Hamiltonian 
\begin{widetext}
\begin{eqnarray}
	\hat{\mathcal{H}}_\text{LOW}\left(\bm{k},\theta\right)&=&
	-\left(12J+24J^\prime\right)a^\dagger_{\bm{k}}a_{\bm{k}}
	-3\varphi_{\bm{k}}^\prime\left(a_{\bm{k}}a_{-\bm{k}}+a_{-\bm{k}}^\dagger a_{\bm{k}}^\dagger\right)
	+\left[\varphi_{\bm{k}}\left(\theta\right)+2\varphi_{\bm{k}}^\prime\right]
	\left(a_{-\bm{k}}^\dagger+a_{\bm{k}}\right)\left(a_{\bm{k}}^\dagger+a_{-\bm{k}}\right) \\
	&=& 
	\left[a^\dagger_{\bm{k}},a_{-\bm{k}}\right]
\left[
\begin{matrix}
	\varphi_{\bm{k}}(\theta)+2\varphi^\prime_{\bm{k}}-6J-12J^\prime & \varphi_{\bm{k}}(\theta)-\varphi^\prime_{\bm{k}} \\
	\varphi_{\bm{k}}(\theta)-\varphi^\prime_{\bm{k}}  & \varphi_{\bm{k}}(\theta)+2\varphi^\prime_{\bm{k}}-6J-12J^\prime
\end{matrix}
\right]	
\left[
\begin{matrix}
	a_{\bm{k}} \\
	a^\dagger_{-\bm{k}}  
\end{matrix}
\right]	
+N\left(6J+12J^\prime\right)
\end{eqnarray}
\end{widetext}
with the auxiliary functions 
\begin{subequations}
	\begin{eqnarray}
		\varphi_{\bm{k}}\left(\theta\right)&=&2J\sum_\gamma\sin^2[2\theta_\gamma+\theta]\cos[\bm{k}\cdot\bm{a}_\gamma],\\
		\varphi_{\bm{k}}^\prime&=&4J^\prime\cos[\bm{k}\cdot\bm{a}_\gamma].
	\end{eqnarray}
\end{subequations}
It can be diagonalized via the Bogoliubov transformation~\cite{BLAIZOT86}
\begin{eqnarray}
	\left[
	\begin{matrix}
		a_{\bm{k}} \\
		a^\dagger_{-\bm{k}}  
	\end{matrix}
	\right]	
	=
	T_{\bm{k}}
	\left[
	\begin{matrix}
		b_{\bm{k}} \\
		b^\dagger_{-\bm{k}}  
	\end{matrix}
\right], 		
\end{eqnarray}
which relates the Holstein-Primakoff bosons with orbital excitation modes. 
Notably, the bosonic statistics require that $T_{\bm{k}}$ satisfies the para-unitary condition
\begin{eqnarray}
	T^\dagger_{\bm{k}}\sigma_zT_{\bm{k}}=\sigma_z.
\end{eqnarray} 
Therefore, the linear orbital wave Hamiltonian can be written in terms of diagonalized bosons 
\begin{eqnarray}
	H_\text{LOW}\left(\theta\right)=\sum_{\bm{k}}\omega_{\bm{k}}(\theta)b^\dagger_{\bm{k}}b_{\bm{k}}
	+NE_{\text{ZP}}\left(\theta\right)
\end{eqnarray}
where the orbital wave excitation is 
\begin{eqnarray}
	\omega_{\bm{k}}(\theta)= 2\sqrt{[\varphi_{\bm{k}}(\theta)+2\varphi^\prime_{\bm{k}}-6J-12J^\prime]^2
		-[\varphi_{\bm{k}}(\theta)+\varphi_{\bm{k}}^\prime]^2}\nonumber
\end{eqnarray}
and the energy from zero-point motion per site is 
\begin{eqnarray}
	E_{\text{ZP}}\left(\theta\right)=\frac{1}{2N}\sum_{\bm{k}}\omega_{\bm{k}}(\theta)+6J+12J^\prime.
\end{eqnarray}
Finally, the zero-point energy can be evaluated numerically.


%

\end{document}